\documentclass[preprint2,12pt]{aastex}
\def\simless{\mathbin{\lower 3pt\hbox
     {$\rlap{\raise 5pt\hbox{$\char'074$}}\mathchar"7218$}}}   
\def\simmore{\mathbin{\lower 3pt\hbox
     {$\rlap{\raise 5pt\hbox{$\char'076$}}\mathchar"7218$}}}   
\def\msun{{\rm M}_\odot}                                
\def\4u{4U 1543--47}                                     

\received{}
\accepted{}
\slugcomment{Submitted to ApJ, xxx 2005}

\lefthead{et al.}
\righthead{\4u}

\begin{document}

\title{Fourier resolved spectroscopy of \4u during the 2002 outburst}

\author{P. Reig and I. E. Papadakis}
\affil{IESL, Foundation for Research and Technology, 
711 10 Heraklion, Crete, Greece}
\affil{Physics Department, University of 
Crete, PO Box 2208, 710 03 Heraklion, Crete, Greece}
\email{pau@physics.uoc.gr,jhep@physics.uoc.gr}

\and

\author{C. R. Shrader and D. Kazanas}
\affil{Exploration of the Universe Division, NASA Goddard Space
Flight Center, Greenbelt, MD 20771}
\email{Chris.R.Shrader@gsfc.nasa.gov,Demos.Kazanas-1@nasa.gov.}


\begin{abstract}

We have obtained Fourier-resolved spectra of the black-hole binary \4u in
the canonical states (high/soft, very high, intermediate and low/hard)
observed in this source during the decay of an outburst that took place in
2002. Our objective is to investigate the variability of the spectral
components generally used to describe the energy spectra of black-hole
systems, namely a disk component, a power-law component attributed to 
Comptonization by a hot corona and the contribution of the iron line due
to reprocessing of the high energy ($E \simmore 7$ keV) radiation.  We
find that {\em i)} the disk component is not variable on time scales
shorter than $\sim$ 100 seconds, {\em ii)} the reprocessing emission as
manifest by the variability of the Fe K$\alpha$ line responds to the
primary radiation variations down  to time scales of $\sim 70$ ms in the
high and very-high states, but longer than 2 s in the low state, {\em
iii)} the low-frequency QPOs are associated with variations of the X-ray
power law  spectral component and not to the disk component and {\em iv)}
the spectra corresponding to the highest Fourier frequency are the hardest
 (show the flatter spectra) at a given spectral state. These results 
question models that explain the observed power spectra as due to 
modulations of the accretion rate alone, as such models do not
provide any apparent reason for a Fourier frequency dependence of 
the power law spectral indices.


\end{abstract}


\section{Introduction}

\4u\ belongs to the group of black-hole X-ray novae \citep{tana96,cher00}.
These  are transient X-ray binaries in which the compact companion is a black
hole and the optical companion a late-type star. They owe their name to the
fact that they occasionally exhibit a large increase of their  X-ray luminosity
(i.e. outbursts), presumably due to a sudden increase of  the mass accretion
rate onto the black hole. At the peak of the outburst the X-ray luminosity may
reach the Eddington limit. In \4u\ these outbursts  are recurrent with a
quasiperiod of 10-12 years. Previous outbursts have been observed in 1971
\citep{mati72}, 1983 \citep{kita84,woer89}, 1992 \citep{harm92} and 2002
\citep{park04,kale05}.

The first report on the observation of the optical counterpart to \4u\ was
given by \citet{pede83}. \citet{oros98} measured the radial velocity curve
of the system and derived a mass function $f(M)=0.22\pm0.02$, an orbital
period of $P_{\rm orb}=1.123\pm0.008$ days and estimated the distance to
be 9.1$\pm$0.1 kpc. They argued that if the secondary star has a mass near
the main sequence values for early A stars \citep{chev92} then the mass of
the primary must be in the range 2.7-7.5 $\msun$. Recently, \citet{park04}
gave a value of $9.4\pm 2 \msun$ (based on work in preparation by J.
Orosz). Thus  \4u\ is very likely to contain a black hole.

Additional evidence of the presence of a black hole in \4u\  is provided
by the  specific sequence of X-ray spectral states  that the system
follows as its  outbursts evolve, usually associated with  accretion onto
black holes.  At  the peak of its recent outburst, while the $2-10$ keV
flux was a large fraction of the Eddington luminosity, the source
exhibited a soft, thermally dominated spectrum and little variability,
i.e. it was  in the High/Soft or thermally dominated  state (HS;
\citet{klis05,mcli04}). As the flux decreased, the source entered the Very
High or steep power-law dominated state (VHS), characterized by broad-band
variability, increased contribution  of the power-law flux and the
presence of QPOs. At the end of the VHS the source showed a sharp increase
in the power-law flux and rms amplitude of variability without a noticeable
change in the photon index. According to \citet{kale05}, \4u\ entered the
intermediate state (IS). Just before the quiescent  state the source went
through the Low/Hard state (LS).  At this state the thermal component was
almost absent, the spectrum was dominated by the power-law,  and the power
spectrum displayed a simple broken power-law shape with rms of
$\sim$20-30\%. 

The X-ray spectral and timing evolution of the source during its 2002 outburst
has been studied in detail by \citet{park04} (HS, VHS and IS), \citet{kale05}
(IS and LS) and \citet{palo05} (quiescent state). The outburst started around
June 15, 2002, and lasted for over one and a half months.  \citet{park04} used
49 RXTE observations that were obtained during the first 35 days of the
outburst,  while \citet{kale05} used 39 observations that were taken $\sim 25$
days after the onset of the outburst. 

In the present work we use archival RXTE data collected at various  epochs
during the latest outburst of \4u. Our aim is to study its Fourier
resolved spectra at various frequency bands during the different spectral
states which the source attains during the  evolution of the outburst.
This latter fact provides the opportunity of studying the variability
properties of the different spectral components of accreting compact
objects (i.e. power law, disk emission and iron line) in different
spectral states of the same object, thus eliminating the ambiguities of
referring to specific states at different objects with different masses
and different Eddington ratios at specific luminosities. 

Our study explores the variability properties on short time scales ($\sim 100$
sec) compared to those of recent studies  \citep{park04,kale05} that
investigated the variability properties of the  above spectral components on
time scales of $\sim 1$ day (the typical  time interval between successive RXTE
observations). Our work follows  the lines of a similar study by \citet{revn99}
and \citet{gilf00} who explored the  spectral variability of Cyg X-1 over
similar time scales ($\sim 0.01 - 100$  sec) during the different spectral
states of the source. A similar approach was also used by \citet{papa05},
who studied the  spectra of the AGN MCG 6-30-15.

While we examine the variability properties of the entire spectrum, we  pay
particular emphasis on those of the Fe K$\alpha$ line; the variation of  this
feature, due almost exclusively to the reprocessing of the harder X-ray 
radiation, is most sensitive to the  geometry of reprocessing matter  in the
vicinity of the accreting object and it can be used to infer its structure. 

In \S 2 we outline the details of our observations and data reduction 
procedure, while in \S 3 we present the results of our analysis. In  \S 4 we
discuss in detail the results of the variability of each  spectral component
and we comment on its significance and implications on the dynamics and
geometry of the accreting matter while in \S 5 we outline our general
conclusions. 

\section{Observations and Data Reduction}

We obtained a total of approximately 60 ksec of data from the RXTE 
archives spanning June 18 to August 4, 2002, thus sampling the outburst 
from near its peak to well into the late decline stages. Typical count
rates (PCA instrument) ranged $\sim 10^4$ near outburst peak to $\sim10^2$
at the late decline phase. All the data  were obtained from RXTE program
IDs P70133 and P70124, the latter  covering the decline phase of the
outburst. 

Data recording and packing in RXTE can be done in many different ways
depending on the brightness of the source and the spectral and timing
resolution requested. The specific observational modes are selected by the
observer and may change during the overall observation. In order to ensure
homogeneity in the reduction process the energy resolution was restricted
to be 16 channels covering the energy band 2-15 keV, as this configuration
could be achieved during the entire duration of the outburst.

\begin{figure}
\epsscale{1}
\plotone{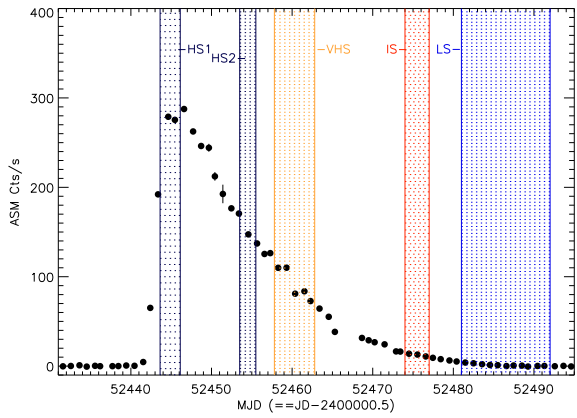}
\caption{Plot of the $2-12$ keV ASM light curve. The shaded boxes indicate the
time periods that we selected in order to extract data from the pointed RXTE
observations within these periods.}
\label{lc}
\end{figure}

Fig.~\ref{lc} shows a plot of the daily average, $2-12$ keV, ASM light
curve of the source during its outburst. For the purposes of the present
work we have selected five observing intervals corresponding to four
different spectral  states of the source, according to the classification
of \citet{park04} and \citet{kale05}. These are shown with the shaded boxes in
Fig.~\ref{lc}. 

The first interval includes the two time ranges MJD  52443.7-52446.1 and
MJD 52453.5-52455.1 (which we refer to as HS1 and HS2, respectively), 
during which the source was in its HS. The total on-source times
were $\sim 18.4$ ksec and $\sim 7.6$ ks respectively. Note that the HS2
period is rather close to the chosen VHS time interval, while the HS1
period covers part of the rise. In this way we can investigate possible
differences in the variability behavior of the source while in its high
state. The second interval spans MJD 52457.8-52460.5 and corresponds to a
period when the source was at its VHS.  It contained 6 observations,
amounting to 10.9 ks. We also  considered four observations  between MJD
52474.2 and 52477.2, which correspond to the IS between the VHS and the LS.
Note, however, that the three observing intervals on July 21 2002 (MJD
52476;  program ID  P70132) were not included in our analysis because a
different onboard spectral (and time) binning was used. Finally, ten
observations between MJD 52481.1 and 52491.5 which corresponds to the LS of
the source. The observing time for the IS and LS were 6.9 and 13.3 ksec,
respectively.

Light curves were extracted for each onboard channel range using the 
current RXTE
software\footnote{http://heasarc.gsfc.nasa.gov/docs/software/lheasoft/},
binned at a resolution of 0.015625 s. We then divided the data into
128-s segments and, following the prescription of \citet{revn99}, we
obtained the Fourier resolved spectra of the source in the following broad
frequency bands:  0.008-0.5 Hz, 0.5-5 Hz and 5-15 Hz. 

Contrary to typical temporal studies which provide the Power Spectral
Densities (PSD) of the source in an entire energy band, Fourier resolved
spectra  provides instead the source spectra at different Fourier frequency
ranges. This method consists of producing the PSD for every energy bin of the
spectrum (i.e. the energy-dependent rms amplitude) and then weighing each
bin in the energy spectrum with the corresponding rms amplitude.

\subsection{Energy spectral analysis}

\begin{figure}
\epsscale{1.}
\plotone{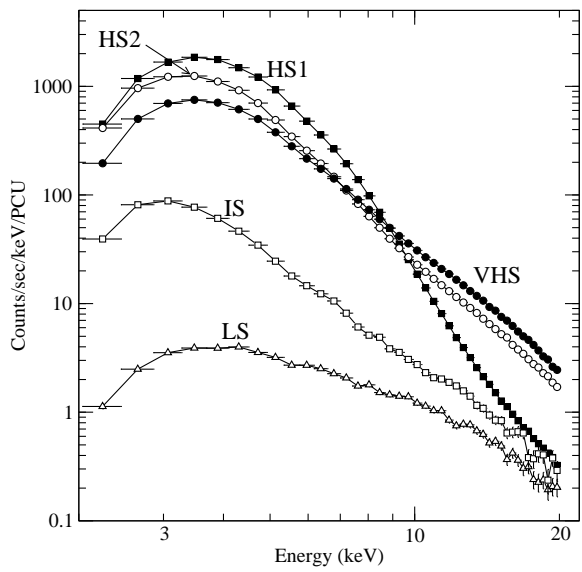}
\caption{Plot of the $2-20$ keV energy spectra of the source during the various
spectral states. The spectra have been labeled accordingly, and are  divided by
the number of the PCU units that were operated in each period.}
\label{sp}
\end{figure}

Fig.~\ref{sp} shows typical $2-20$ keV energy spectra of the source
corresponding to the observational periods selected. The spectra were
extracted from PCA {\em Standard 2} mode data. The response matrix and
background models were created using the standard HEADAS  software, version
5.3. The number of detectors (PCU) that were switched on varied
for each observation, and, in order to be able to compare the
spectra, they were divided by the respective number of PCUs.

The filled squares and open circles in Fig.~\ref{sp} show the spectrum of
the source in the HS (HS1 and HS2, respectively), using data from the
observations June 19 (when the source reached its maximum flux) and June
28, 2002 \citep[Obs. No. 5 and 16 in][]{park04}. Open circles and open
squares in the same figure show representative spectra of the source
during the VHS and IS, respectively, using data from the observations
performed in July 4 and 19, 2002 \citep[Obs. No. 22 and 44 in][]{park04}.
Finally, the open triangles show a representative LS spectrum, using data
taken in August 1, 2002 \citep[Obs. No. 19B in][]{kale05}. The  spectral
evolution with time is apparent in this figure.  The HS spectra are
characterized by a dominant thermal black body and a weak, steep power law
component. The power-law flux increases during the VHS, and becomes the
dominant component in the LS. As the total flux of the source decreases,
the power-law slope becomes harder.  

Although the spectral evolution of the source has been studied extensively
in the past, we fitted the spectra using the same model components as in
\citet{park04, kale05}, i.e. a {\tt wabs} model, to take account of the
interstellar absorption effects (with $N_H=4.1 \times 10^{21}$ cm$^{-2}$
that was kept fixed, \citet{park04}), a multicolor blackbody accretion
disk model \citep{mits84,maki86}, a power-law model, a narrow Gaussian
line  (i.e. $\sigma$ fixed at  0.1 keV, smaller than the spectral
resolution of RXTE) to account for the iron K$\alpha$ line emission, and a
broad smeared absorption edge model ({\tt smedge} in {\tt XSPEC}, with the
width fixed at 7 keV, like in \citet{park04}). The main reason to analyze
the energy spectra is to reduce the spectral resolution of the {\em
Standard 2} spectra to match that of our Fourier-resolved spectra (i.e. 16
bins, in the  2-15 keV band, compared to $\sim$ 50 of the {\em Standard 2}
mode). This is a necessary step in order to be able to compare the results
from the model fitting of the energy spectra with those from the model
fitting of the Fourier-resolved spectra (presented in the following
section). 

The spectral analysis was performed using XSPEC version 11.3.1. We have
added systematic errors of 1\% to all channels and have restricted our
analysis to the 2-15 keV band only (to match the energy band used in the
case of the Fourier-resolve spectra). Our results are listed in
Table~\ref{enespec}.
The errors quoted for the best fit values correspond to the 90\%
confidence limit for one interesting parameter. In the case when the error
is large enough and the best-fit parameter value is consistent with being
zero, we simply note the best-fit value plus  upper error, and we accept
it as the upper 90\% limit for the respective parameter. 

Our results are entirely consistent with those reported by  \citet{park04,
kale05} for the respective observations. In the case of the HS, VHS and IS
spectra, our best-fit estimate of the  equivalent width (EW) of the iron
emission line is systematically smaller than that reported by \citet{park04},
the main reason being the use of a narrow Gaussian line model in our case
(which fits well the reduced resolution spectra that we are using). 

Note that in the case of the LS spectrum, the addition of a narrow Gaussian
line at $\sim 6.4$ keV to the single power-law fit  reduces the $\chi^{2}$ 
by 9.7 (for two additional parameters), which is significant at the 91\%
level.

\section{Fourier resolved spectral analysis}

The usefulness of the frequency-resolved spectra lies on the fact that
they  receive significant contribution only from the spectral components
that  are variable on the time scales sampled by the observations.
Therefore by  performing Fourier-resolved spectroscopy we can investigate
whether the  various spectral components in the overall spectrum of the
source (i.e.  disk black-body, power-law, iron line) are variable at each
frequency  range considered.  In general, the interpretation of the
Fourier resolved spectra is not unique and requires additional
assumption about the cause of variability. However, for
the case of the iron Fe K$\alpha$ line and the Compton reflection
components which are thought to result from the reprocessing of higher
energy ($E \ge 7$ keV) radiation and are filtered by the well understood
light travel-time effects, the Fourier-resolved analysis can provide
meaningful constraints on the geometry of reprocessing  matter with
respect to the source of the hard radiation.

In this section we present the results of our Fourier spectral fitting
analysis for each spectral state and compare the resulting best-fit parameters
to those of the previous section (i.e, those obtained from the average energy
spectra).   As before, {\tt XSPEC} version 11.3.1
was used for the model fitting. Most spectral fits yielded residuals
attributable to absorption from the Xenon L edge at 4.78 keV. In order to
account for this instrumental feature, we included in all model fits a
Gaussian line model with central peak at 4.5--5 keV and fixed width
($\sigma=0.1$). We have also added, in all cases, an absorption component
($N_H=4.1 \times 10^{21}$ cm$^{-2}$). A uniform systematic error of 1\% was
added quadratically to the statistical error of all Fourier spectra in each
energy channel. Errors quoted for the best-fit values correspond to the
90\% confidence limit for one interesting parameter (or to 90\% upper
limits in the case when the errors are too large). We describe the results
of our analysis for each of the source's spectral states below.

\subsection{The High/Soft State}

We first fitted the HS1 and HS2 Fourier spectra with a simple power-law 
model. We found that this model did not provide an acceptable fit to any of
the  Fourier spectra (note that, due to the low variability amplitude
during the high state, we could not estimate a high frequency Fourier
spectrum for the HS1 data). In both cases,  the residuals reveal the
presence of an emission and absorption feature at $\sim 6-7$ and $\sim 7-9$
keV, respectively. We then fitted the Fourier spectra with a model that
consists of a power law, a Gaussian line and an edge (a simple {\em edge}
model provided a better fit than the  {\em smedge} that was used in the
case of the energy spectrum).  The energy and depth of the edge and the
line energy were allowed to float as free parameters, but were forced to be
identical in all three Fourier spectra. The line normalization was allowed
to float in the three spectra. 

\begin{figure*}
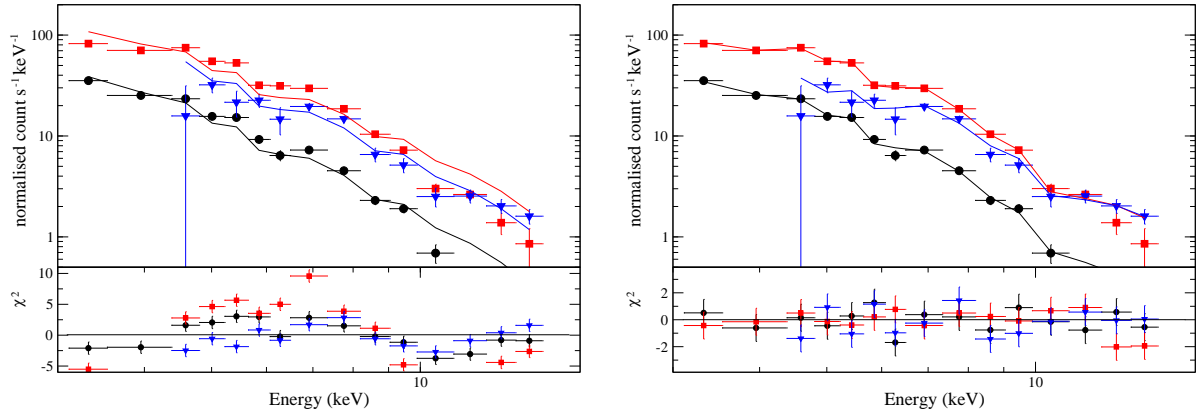

\epsscale{1.}
\begin{tabular}{cc}
\plotone{f3a.eps} &
\plotone{f3b.eps} 
\end{tabular}
\caption{Energy spectra of the high state (HS) fitted to an absorbed
power-law (left) and an absorbed power-law plus an iron line and edge
(right). Different symbols represent different frequency ranges: 0.008-0.5
Hz (circles), 0.5-5 Hz (squares) and 5-15 Hz (triangles).
\label{hs}}
\end{figure*}

The best-fit parameter values for this model in the case of the HS1/HS2
spectra are listed in Table~\ref{simulfit}. Since they are consistent
within the errors, we combined the individual HS1 and HS2 Fourier spectra
and estimated the overall HS Fourier spectra. A simple power law model does
not fit the data well ($\chi^2$ of 517 for 37 dof). The left panel in
Fig.~\ref{hs} shows the overall HS Fourier spectra, with the best-fit 
power-law model and the model residuals. The residuals clearly indicate the
presence of a line emission and absorption edge features in the spectra.
When these components are added to the model (see Fig.~\ref{hs}, right
panel) the fit improves considerably ($\chi^2=31$ for 24 dof). The best-fit
parameter values for this model are also listed in Table~\ref{simulfit}. 

When the presence of the iron line is significant the equivalent width
measured in the frequency-resolved spectra is larger than that found in the
corresponding HS energy spectrum (listed in Table~\ref{enespec}).
Similarly,  the best-fit edge energy of $\sim 9$  keV (a value
representative of material with a high degree of ionization) appears to be
significantly higher than the  estimate of $7.5-8$ keV, reported in
Table~\ref{enespec}. 

As for the best-fit power-law slopes, we observe a significant hardening
with increasing frequency. Compared to the overall spectral slope of $\sim
2.5$ that characterizes the power-law component in the high state
\citep{park04}, the low- and medium-frequency spectra are  significantly
steeper, while the high-frequency Fourier-spectrum slope is consistent with
it. 

\subsection{The Very High State}

\begin{figure*}
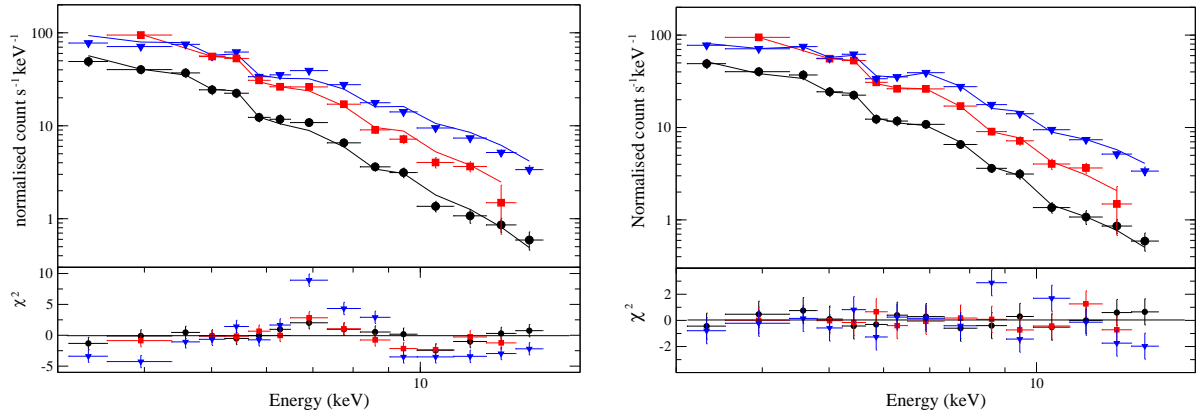

\epsscale{1.}
\begin{tabular}{cc}
\plotone{f4a.eps} &
\plotone{f4b.eps} 
\end{tabular}
\caption{Energy spectra of the very high state (VHS) fitted to an absorbed power-law
(left) and an absorbed power-law plus an iron line and edge (right).
Symbols as in Fig.~\ref{hs}.
\label{vhs}}
\end{figure*}

The left panel of Fig.~\ref{vhs} shows the best-fit power-law model to the
VHS Fourier spectra. As with the HS spectra, it does not fit the data well
($\chi^2=210$ for 33 dof). Significant emission and absorption features
appear at energies above $\sim 5-6$ keV. The right panel in Fig.~\ref{vhs}
shows the three VHS frequency-resolved spectra with the best-fit 
"power-law+Gaussian+absorption edge" model, which does provide a
significantly better fit to the data ($\chi^2=29$ for 24 dof). The
best-fit parameter values are listed in Table~\ref{simulfit}. 

The iron line is clearly detected in the medium- and high-frequency
spectra. Interestingly, the line energy is larger than the corresponding
value in both the HS Fourier-spectra, and the VHS energy spectrum. The
best-fit edge energy value is also larger than that in the VHS energy
spectrum. In contrast, the edge optical depth in the VHS Fourier spectra
appears to be smaller than in the VHS energy spectrum. 

The power-law slope becomes harder as the frequency of the Fourier spectra
increases. Compared to the overall power-law spectral slope, the low- and
medium-frequency values are steeper by $\Delta\Gamma\sim 0.9-1$, while the
high-frequency spectral slope is harder by $\Delta\Gamma\sim 0.2$.

\subsection{The Intermediate and Low States}

\begin{figure*}
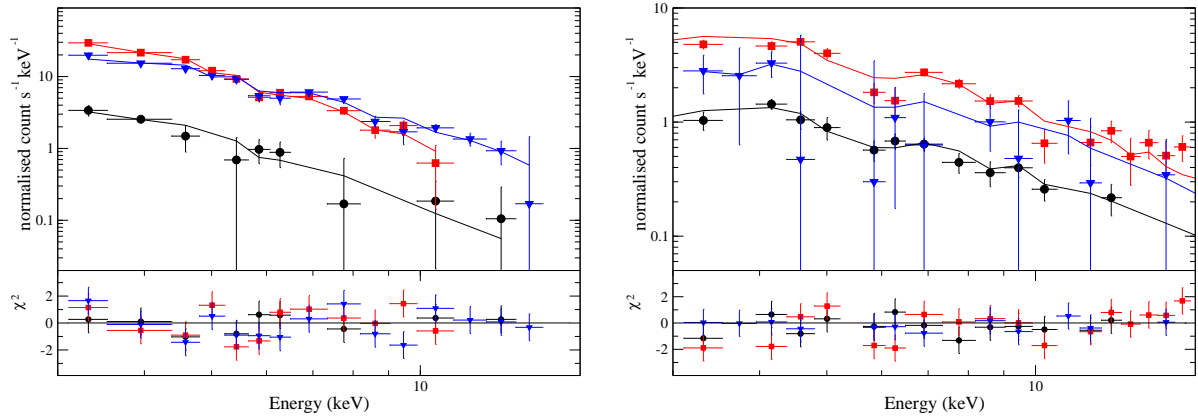

\epsscale{1.}
\begin{tabular}{cc}
\plotone{f5a.eps} &
\plotone{f5b.eps} 
\end{tabular}
\caption{Energy spectra of the intermediate state (left) and hard/low
state (right) fitted to an absorbed power-law model.
Symbols as in Fig.~\ref{hs}.
\label{lhis}}
\end{figure*}

The right and left panels in Fig.~\ref{lhis} show  the best power-law
model fits to the IS and LS Fourier spectra. In these cases,  the model
provides an acceptable fit ($\chi^2=30$ for 30 dof, and 32.6 for 24 dof,
respectively). During the  IS and LS observations the strength of the
power law component increased, while its slope flattened reaching
$\Gamma\sim 1.7$ during the LS period.  Although the iron line and
absorption edge are still present in the IS energy spectrum of the source,
and the line may also be detectable in the energy spectrum during the LS
(Table~\ref{enespec}), these features are no longer evident in the Fourier
resolved spectra, in marked contrast with the Fourier resolved spectra of
the HS and VHS described above. 

Looking at the residuals in the left panel of  Fig.~\ref{lhis}, one can
see the same structure as in the HS and VHS, namely, the characteristic
"wiggle" in the 5-12 keV energy range (i.e., an excess of flux at about
$6-8$ keV and a deficit at about 9 keV). For that reason we added a
Gaussian line component in the IS  model spectrum and we repeated the
model fitting, keeping the line energy fixed to 6.4 keV. However, the
quality of the fit did not improve significantly. We conclude that, if
present, the strength of the  iron line and the absorption edge must be
significantly decreased, compared to that of the same features  in the HS
and VHS Fourier-spectra. 

The spectral slope of the IS Fourier spectra is significantly steeper than
the power-law slope in the respective  Fourier spectra of the HS and VHS 
(by a factor of $\Delta \Gamma \sim 0.5-1$).  The low- and medium-frequency
IS Fourier spectra are steeper than the overall spectrum by
$\Delta\Gamma\sim 1.5$. The high-frequency slope is flatter, but still
steeper than the overall spectrum by  $\Delta\Gamma\sim 0.6$. Finally, in
the LS  case, the flux of the source is too low to obtain a meaningful
high-frequency (5--15 Hz) Fourier spectrum. The low- and medium-frequency
spectra are slightly steeper than the overall energy spectrum 
($\Delta\Gamma\sim 0.3-0.5$).

\section{Discussion}

In the previous sections we discussed our analysis and 
results of the Fourier Resolved Spectroscopy of 4U 1543-73 during the
entire evolution of its 2002 outburst. The main goal of our work 
is to  enlarge the sample of objects
analyzed  in this specific way,  in an attempt to uncover and  establish
systematics  associated with their spectro-temporal properties, different
from the  usual ones provided by  simply their power spectral density
(PSD).  We found that a single 
power-law component does not provide good fits to most of the 
Fourier-resolved spectra and that the signatures typical 
of X-ray radiation reprocessing (such as iron line and edge) were
required in order to obtain acceptable fits. In addition, 
no disk component was found in the Fourier-resolved spectra whose 
hardening with increasing frequency appears to
be a general characteristic in all states.  In this section we discuss the
implications of these results and investigate the temporal properties of
the model components that are generally used in black-hole spectral
analysis. 


\subsection{The disk component}

One of the most striking results of our analysis is the  absence of
variability in the multicolor blackbody disk component that provides the
dominant flux in the observed HS and VHS spectra. The absence of
variability in this component is manifest  by the fact that  the
Fourier-resolved spectra are well fitted by a power law component only
(plus an iron line and edge) when the system is in the HS and VHS.  The
fact that the contribution of the disk multicolor blackbody component is
negligible in all Fourier-resolved spectra  suggests  that  the disk is
not variable on time scales shorter than $\sim$ 100 s. Even in the HS,
when the disk is believed to extend down to the last stable orbit (and the
variability time scales could indeed be short), the disk component is not
required in the Fourier-resolved spectra at any of the frequency bands we
examined. 

This absence of variability is consistent with the magnitude of the 
viscous time scale of a disk with temperature $\simeq 1$ keV, estimated 
to be 
\begin{eqnarray*}
\tau_{\rm vis} & \simeq & \frac{R_S}{c} x^{5/2} \frac{m_pc^2}{kT} \alpha^{-1}  \\
&\simeq & 1.5 \times 10^3 \left(\frac{M}{10 M_{\odot}}\right)
\left(\frac{T}{1\; {\rm keV}}\right)^{-1} \alpha^{-1} \, {\rm sec}
\end{eqnarray*}

\noindent where $x \simeq 3$ is the disk inner edge radius in units of the
Schwarzschild radius ($R_S$) and $m_p$ is the proton mass. We conclude
therefore that although the disk blackbody flux changes from day to day
\citep[see Fig.~2 in][]{park04}, the disk is stable on much shorter time
scales (i.e. less than 100 sec). A similar result was also reported in the
case of Cyg X--1 \citep{gilf00}, who found that the Fourier-resolved
spectra of Cyg X--1 too are well fitted by a simple power-law component at
all frequency bins, with no indication for a multi-color disk component,
both when the system is in the High and Low States. The absence  of
rapid variability of the disk component, a result already noted by 
\citet{miya94}, appears to be a general trend in this class of objects.

This lack of disk variability puts constraints on models that attempt to
attribute the observed PSDs  as due to modulation of the accretion rate onto
the compact object alone. At a minimum one would expect some variability of
this component due to the reprocessing of the variable X-ray component on the
disk and its re-emission as disk radiation. However, since the variable power
law component represents only a small fraction of the disk luminosity, such 
variations, while  presumably present, are hard to discern 
because of their small
amplitude. For the same reason, we also believe that, at least in HS1,
variations intrinsic to the disk, necessary for the dissipation of its kinetic
energy, are too small to cause significant variations in its flux over the
sampled time scales.

\subsection{The iron line and edge}

The results presented in \S 2.1 indicate the presence of a fluorescent
K$\alpha$ iron line  at $\sim$ 6.4 keV and an edge at $\sim 7-7.5$ keV in
the HS, VHS and IS energy spectra, in agreement with the results of
\citet{park04}.  These features constitute the main signatures for
reflection in cold material of the primary source of X-rays. Our results
presented in \S 3 exhibit the presence of similar features in the Fourier
spectra for all frequencies (even the highest), when the source is in the
HS and VHS (the fact that the line is not clearly detected in the
low-frequency Fourier spectrum is almost certainly due to the fact that
this spectrum has the lowest signal-to-noise among the three Fourier
spectra). The Fourier-resolved spectra in the IS and  LS are well fitted by
a single power law model only. This implies that either the reflection
features are absent  or, if present, their strength must be significantly
reduced when the system is in these states.

In the case of a conventional (i.e. non-Fourier resolved) energy spectrum, the
equivalent width  of the iron line (assuming that the reprocessing matter is 
neutral) is proportional to the ratio of the reflection component amplitude to
that of primary radiation. However, the equivalent width of the iron line
determined from a Fourier frequency resolved spectrum corresponds to the ratio
of the rms variability amplitude of the reflected component to that of  the
primary emission variations in a given Fourier frequency range $\Delta \nu$,
i.e., to the solid angle of the X-ray source  subtended by reprocessing surface
up to a length scale $L \simless c/ \Delta \nu$. 

Our results show that the equivalent width of the line is $\sim 250-450$ eV at
all frequency bins when the system is in the HS and VHS. This suggests that the
reflected emission is fully  responding to the primary radiation variations up
to frequencies $\sim 15$ Hz, or to  to time scales of $\simeq 70$ ms.  On the
other hand, our results also show that, when the system is in the IS and LS,
the reflected component does not follow the primary emission variations on time
scales shorter than 2 s (and perhaps even longer). 

\citet{gilf00} reported similar results for Cyg X--1, when the source was in
its HS and LS. They suggested that  the most straightforward explanation  of
their results (and hence of ours as well) is in terms of a finite 
light-crossing time to the distance of the reflector. The equivalent  width of
the line in the Fourier-resolved spectra should remain roughly constant up to a
frequency which corresponds to the inverse of the light travel time between the
hard X-ray emitting corona and the inner radius of the disk that can reprocess
the X-ray radiation into Fe K$\alpha$.  Using Fig.~6 in \citet{gilf00}, and the
fact that the line equivalent width remains roughly constant up to frequencies
$\sim 15$ Hz when the system is in the HS and VHS, we conclude that the
innermost radius of the reflective material could be as low as $\sim 10 R_{g}$
for a 10 M$_{\odot}$ black hole. The decrease of the line equivalent width in
the IS and LS is consistent with the assumption that the accretion disk does
not extend to small radii any longer. 

In addition to the light travel time effect on the response of the iron line,
the latter can be also influenced by the ionization sate of the reprocessing
medium. As the latter increases, the energy of the iron line and associated
edge increases \citep[][and references therein]{geor91}. The larger values of
the line energy in the VHS Fourier-spectra, $\sim 6.8$ keV,  with respect to
the VHS energy spectrum and the HS Fourier-spectra  ($\sim 6.4$ keV in both
cases) could then imply a higher degree of ionization in the innermost parts of
the disk when the system is in the VHS. This is perhaps expected, since the
power law component, which presumably illuminates the disk, has a larger
luminosity during the VHS. Eventually, for sufficiently large values of the
X-ray flux (more correctly of the ionization parameter) the reprocessing medium
becomes highly ionized, resulting in suppression of both the Fe K$\alpha$ line
and the Compton reflection features \citep{naya00}.

\subsection{The QPO}

During the time interval MJD 52456--52461 \cite{park04} reported the detection
of a QPO. The central frequency of the QPO varied in the range 7--10 Hz and the
Q (coherence) parameter in 5-9 (i.e, a FWHM of $\simless$ 2 Hz). However, when
obtaining the average over the entire period when the QPO is present, the QPO
extends over a wider frequency interval. In fact, we chose the high-frequency
interval, namely 5--15 Hz, so as to cover all the frequency range of the QPO
that appeared when the system was in the VHS. Therefore, the High-Frequency
Fourier-resolved spectrum when the system is in the VHS should be
representative of the energy spectrum of the variability components that
``produce" the QPO in the system.  The fact that no disk component is
statistically required to fit this spectrum,  implies that the QPO is not
associated with the disk emission. 

This result is in accordance with the behavior seen in most black hole systems
\citep[see e.g.][]{swan01,mcli04,klis05}, namely that the QPO usually appears
when the flux is dominated by the hard power-law component (there are no
detected QPOs in the HS). The association between the QPO and the hard
power-law is substantiated by the fact that the QPO amplitude increases with
photon energy, when energy bands beyond the characteristic energy range of a
multicolor blackbody with $kT\sim 1$ keV are considered  \citep{bell97,morg97}.
\citet{muno99} found that when the QPO is present, the power-law flux is much
more variable than the disk flux. Only when the QPO is absent, the blackbody
component is much more variable than the power law. In the case of \4u,
\citet{kale05} show that the QPO frequency does depend on the power law slope,
and decreases with decreasing $\Gamma$, a correlation found to be generally 
present in accretion powered sources \citep{titfior04}.

\subsection{The power-law component}

A common effect seen in all three spectral states is the hardening of the 
power-law component with increasing frequency.  That is, the X-ray
emission associated with the variability of the shortest time scales is
harder than  that associated with the variability at longer time scales. 
Such a behavior is at first glance inconsistent with a model that
attributes all variations to modulation of the accretion rate in a fashion
that reproduces the observed PSDs. We do not see any obvious
reason for such a behavior in the context of this type of model.  This
type of behavior is consistent with that observed by \citet{revn99} and
\cite{gilf00}  in Cygnus X-1 in its low-hard state. However, as pointed
out in the latter work, the variable power law component  of this source
in its soft state is independent of the Fourier frequency, a fact not in 
complete agreement with the  results of the present work. 

A closer look at the results listed in Table 2 shows that at all states
(except HS1) the high-frequency (HF) spectra are the hardest, while the
power-law index of the low-frequency (LF) spectra is similar to that of the
medium-frequency (MF) spectra, that is,  $\Gamma_{\rm LF}\sim \Gamma_{\rm
MF}$, while $\Gamma_{\rm {LF,MF}} > \Gamma_{\rm HF}$.  In other words, we
observe a rather large $\Delta \Gamma$ jump at frequencies higher than 5
Hz. This is the frequency at which the QPO appears. Furthermore, the 2-30
keV power spectrum during the HS and LS show a slope change at $\sim 5$ Hz
- see upper left and bottom right plots in Fig.8 of \citet{park04}. This
feature is  probably related to the dynamics of accretion, whose
significance we are not able to assess at this point. However, it appears
that the frequency range where the QPO lies presents a characteristic 
scale for this system that provides a demarcation of its spectral and
timing properties.

Finally, we also observe that  $\Gamma_{\rm LS}<\Gamma_{\rm{HS,VHS}}$ in
the low and medium frequency bins. This  correlation is generally 
observed among the different spectral states of accreting black holes, and
is attributed to cooling of the corona temperature with the increased soft
photon flux associated with the HS and VHS. 

\section{Conclusions}

\4u is the third (besides Cygnus X-1 and GX 339-4, \citet{revn01}) source
amongst Galactic Black-Hole binaries for which Fourier resolved spectra have
been calculated. The results from the energy spectral analysis of the three
sources reveal several common characteristics, most importantly different
spectral states characterized by soft and hard components, the former at higher
luminosity than the latter. 

Their Fourier resolved spectra at the corresponding spectral states
exhibit also similarities: a) in the low/hard state the
Fourier spectra tend to be harder with increasing Fourier frequency, while
the Fe K$\alpha$ line is more prominent at lower Fourier frequencies. This
result has been attributed to the size of the accretion disk inner radius,
which may be set at distances $R \simeq 100 R_S$ when the systems are in
their LS. While this explanation can account for the absence of Fe line at
high frequencies, it cannot account for the large value of the Fe line EW
(in the case of Cyg X-1 for example)  if the size of the X-ray emitting
region is $\sim 10R_S$, as it is usually assumed. b)  In the soft spectral
states, the Fe line is present independent of the Fourier frequency. In
the case of \4u, we also  observe a general hardening of the spectra 
with Fourier frequency in these states (different to Cyg X-1). c)  A
common characteristic of all Fourier resolved spectra (in all states)  is
the absence of the multicolor disk component in the Fourier resolved
spectra, indicating that this component, while present, is not variable on
time scales as short as a few hundred seconds. This results is in
agreement with the viscous time scales of such disks.

In the case of \4u, we also find evidence of an increased ionization state of
the reflector when the system is in the VHS, while the absence of the disk
component even in the QPO frequency range during the VHS, implies that the QPO
emission is not associated with the disk emission. 

The FRS technique provides a novel look at the structure of accretion powered
sources. The general trends observed in the to-date analyses suggest common
underlying systematics which are not fully yet understood. We believe that
further analysis and modeling along the same lines for other sources is highly
warranted.

Part of this work was supported by the General Secretariat of Research
and Technology of Greece.

\clearpage
\begin{deluxetable}{lccccc}
\tabletypesize{\scriptsize}
\tablecaption{Results from the model fits to the energy spectra 
$^a$ \label{enespec}}
\tablewidth{0pt}
\tablehead{\colhead{Parameter}&\colhead{HS$_1$}&\colhead{HS$_2$}&\colhead{VHS}&\colhead{IS}&\colhead{LS}}
\startdata
$T_{\rm col}$ (keV)     &$0.99\pm 0.01$      	& $0.88\pm0.02$ 	
&$0.87^{+0.02}_{-0.05}$	&$0.6\pm 0.1$	&$0.35^b$ \\
$\Gamma$		&$3.6\pm 0.2$ 		&$2.6^{+0.2}_{-0.1}$	
&$2.7\pm0.2$	&$2.5\pm 0.1$ 		&$1.73^{+0.04}_{-0.03}$    \\      
E$_{\rm Fe}$ (keV) 	&$6.2\pm 0.2$		&$6.4\pm 0.2$   	
&$6.3^{+0.2}_{-0.3}$  &$6.5^{+0.1}_{-0.2}$    & $6.3^{+0.2}_{-0.4}$  \\     
E$_{\rm edge}$ (keV)	&$7.5\pm 0.1$ 		&$7.8^{+0.7}_{-0.5}$   	
&$7.5^{+0.6}_{-0.5}$  &$7.1^{+0.3}_{-0.0}$    & --     \\	
$\tau$             	&$1.4^{+0.1}_{-0.2}$		&$0.8\pm 0.4$  
&$1.3^{+0.9}_{-0.2}$    & $1.9^{+0.5}_{-1.2}$		  & --    \\   
EW(Fe) (eV)        	&$92\pm 8$		&$91^{+8}_{-11}$	       	
&$90^{+22}_{-17}$              &$210^{+195}_{-120}$	& $<260$    \\	  
$\chi^2_{\rm red}$/dof/prob &1.5/6/0.16 	&0.69/6/0.66 	       
&1.3/6/0.75         & 2.4/6/0.025	      &1.5/9/0.14 \\
\tableline
\enddata
\tablenotetext{a}{Errors and upper limits represent the 90\% confidence limits.}
\tablenotetext{b}{Parameter value kept fixed, as in \citet{kale05}}
\end{deluxetable}

\clearpage

\begin{deluxetable}{lcccccc}
\tabletypesize{\scriptsize}
\tablecaption{Results from the model fits to the Fourier resolved spectra 
$^a$ \label{simulfit}}
\tablewidth{0pt}
\tablehead{\colhead{Parameter}&\colhead{HS$_1$}&\colhead{HS$_2$}&\colhead{HS}&\colhead{VHS}&\colhead{IS}&\colhead{LS}}
\startdata
\tableline
\multicolumn{7}{c}{0.008--0.5 Hz}  \\
\tableline
$\Gamma$		&3.8$^{+0.2}_{-0.2}$ 	&3.2$^{+0.3}_{-0.4}$   &3.6$^{+0.2}_{-0.2}$  &3.6$^{+0.1}_{-0.2}$    &4.0$^{+1.2}_{-0.9}$    &2.2$^{+0.3}_{-0.2}$    \\      
E$_{\rm Fe}$ (keV) 	&6.8$^{+ 0.4}_{-0.6}$	&6.4$^{+0.4}_{-0.2}$   &6.4$^{+0.3}_{-0.2}$  &6.8$^{+0.1}_{-0.2}$    & --		  & --    \\   
norm ($\times 10^{-4}$) &4$^{+3}_{-3}$ 		&$<4.7$		       &$<3.5$  	     &$<7.3$		      & --		  & --    \\  
E$_{\rm edge}$ (keV)	&9.2$^{+0.3}_{-0.4}$ 	&9.0$^{+0.2}_{-0.2}$   &9.0$^{+0.2}_{-0.2}$  &9.1$^{+0.5}_{-0.3}$    & --		  & --     \\	
$\tau$             	&0.8$^{+0.5}_{-0.5}$ 	&$<1.4$		       &1.1$^{+0.5}_{-0.5}$  &0.6$^{+0.4}_{-0.2}$    & --		  & --    \\   
EW(Fe) (eV)        	&$<1000$		&$<1350$	       &$<575$              &$<900$		      & --		  & --    \\	  
\tableline 
\multicolumn{7}{c}{0.5--5 Hz}  \\
\tableline
$\Gamma$       		&3.1$^{+0.3}_{-0.2}$ 	&2.8$^{+0.1}_{-0.1}$   &3.0$^{+0.1}_{-0.1}$  &3.7$^{+0.2}_{-0.2}$    &4.2$^{+0.1}_{-0.1}$    &2.0$^{+0.2}_{-0.1}$ 	\\	
E$_{\rm Fe}$ (keV)	&6.8$^b$	        &6.4$^b$	       &6.4$^b$             &6.8$^b$		     &--		     &--     \\      
norm ($\times 10^{-4}$) &9$^{+7}_{-8}$ 		&13$^{+5}_{-5}$	       &19$^{+6}_{-5}$       &9$^{+6}_{-6}$	     &--		     &--     \\  
E$_{\rm edge}$ (keV)	&9.2$^b$	        &9.0$^b$	       &9.0$^b$              &9.1$^b$		     &--		      &--     \\     
$\tau$         		&1.0$^{+0.7}_{-0.6}$ 	&1.1$^{+0.2}_{-0.2}$   &1.4$^{+0.3}_{-0.3}$  &$<0.56$		     &--		     &--     \\      
EW(Fe) (eV)		&$<1000$	        &450$^{+175}_{-175}$   &435$^{+205}_{-175}$  &270$^{+190}_{-165}$    &--		     &--     \\      
\tableline
\multicolumn{7}{c}{5--15 Hz} \\
\tableline
$\Gamma$       	         &--  			&2.4$^{+0.2}_{-0.2}$   &2.5$^{+0.5}_{-0.7}$   &2.5$^{+0.1}_{-0.1}$    &3.1$^{+0.1}_{-0.1}$    &--     \\      
E$_{\rm Fe}$ (keV)	 &--  			&6.4$^b$	       &6.4$^b$              &6.8$^b$  	      &--		      &--     \\   
norm ($\times 10^{-4}$)  &--  			&10$^{+8}_{-7}$	       &$<25$		       &20$^{+5}_{-5}$	     &--		      &--     \\   
E$_{\rm edge}$ (keV) 	 &--  			&9.0$^b$	       &9.0$^b$               &9.1$^b$  	      &--		      &--      \\   
$\tau$         	         &--  			&1.0$^{+0.4}_{-0.4}$   &1.0$^{+0.4}_{-0.4}$   &0.3$^{+0.1}_{-0.2}$    &--		      &--      \\   
EW(Fe) (eV)    		 &--  			&600$^{+690}_{-390}$   &--		   &390$^{+160}_{-150}$    &--		      &--     \\      
\tableline
$\chi^2_{\rm red}$/dof/prob &0.54/17/0.93 	&1.30/24/0.15 	       &1.22/25/0.21         & 1.21/24/0.22	     & 1.01/30/0.45	     &1.36/24/0.11 \\
\tableline
\enddata
\tablenotetext{a}{Errors represent the 90\% confidence limits. 
Upper limits in the EW of the iron line are also 90\%}
\tablenotetext{b}{Freeze to the first appearance of the parameter}
\end{deluxetable}

\clearpage

\clearpage

\clearpage

\clearpage

\clearpage

\end{document}